\documentstyle[12pt,epsf]{article}
\begin{document}
\setlength{\baselineskip}{7mm}
%
%
%
%
\begin{flushright}
DPNU-97-09, January 1997
\end{flushright}
%
%
%
%
\begin{center}
{\LARGE 
The Phase Structure of the Gross-Neveu Model
}\end{center}\begin{center}
{\LARGE 
with Thirring Interaction at the Next to Leading
}\end{center}\begin{center}
{\LARGE 
Order of $1/N$ Expansion }
\end{center}
%
\begin{center}
{\large Takashi Dateki}\footnote{e-mail: dateki@eken.phys.nagoya-u.ac.jp}
\end{center}
\begin{center}
{\it Department of Physics, Nagoya University \\
Nagoya 464-01, Japan }
\end{center}
%
%
%
%
%
%
%
\begin{abstract}

We study the critical behavior of the $D$ $(2<D<4)$ 
dimensional Gross-Neveu model with a Thirring interaction, 
where a vector-vector type four-fermi interaction is on equal 
terms with a scalar-scalar type one. 

By using inversion method up to the next-to-leading order of $1/N$ 
expansion, we construct a gauge invariant effective potential. 
We show the existence of the chiral order phase transition, 
and determine explicitly the critical surface. 
It is observed that the critical behavior is mainly controlled 
by the Gross-Neveu coupling $g$. 
The critical surface can be divided into two parts by the surface 
$g=1$ which is the critical coupling in the Gross-Neveu model 
at the $1/N$ next-to-leading order, and the form of the critical surface 
is drastically change at $g=1$. 
Comparison with the Schwinger-Dyson(SD) equation is also discussed. 
Our result is almost the same as that derived in the SD equation. 
Especially, in the case of pure Gross-Neveu model, 
we succeed in deriving exactly the same critical line 
as the one derived in the SD equation.

\end{abstract}
%
%
%
%
\section{Introduction}
\indent

One of the most important subject in the elementary particle 
physics is how to treat non-perturbative effect and 
how to obtain non-perturbative information. 
Since the work of Nambu-Jona-Lasinio(NJL)\cite{NJL}, 
chiral symmetry breaking 
have been studied extensively in various models, 
e.g. QCD, strong coupling QED\cite{scqed}, 
gauged NJL\cite{KTY,KMY}, etc.. 
Particularly, the mechanism of dynamical mass generation 
has been applied to models beyond the standard model, 
for example, walking technicolor\cite{TC}, 
top mode standard model\cite{TMSM}, etc.. 
In these models, scalar 4-fermion interaction plays an important role. 
Such a 4-fermion interaction in $D(2<D<4)$ dimensions, 
the Gross-Neveu model\cite{GN}, has been extensively studied. 

Recently dynamical symmetry breaking and dynamical 
mass generation in the Thirring model\cite{Thirring}, 
which has vector-vector type 4-fermion interaction 
in $D$ dimensions$(2<D<4)$, has been studied by many authors 
\cite{Thirring1,iksy,Thirring in inversion}. 
Usually the Thirring model is studied in the form where the fermion 
has a minimal coupling with a massive vector boson through introduction 
of the vector auxiliary field. 
Although this theory (as well as original Thirring model) 
has no local gauge symmetry, several authors treated 
this massive gauge field as a genuine gauge field\cite{Thirring1}. 
Itoh et al. \cite{iksy} have reformulated Thirring model 
as a gauge theory 
by using the hidden local symmetry\cite{hidden local}, 
and emphasized the advantage of maintaining such a manifest 
gauge symmetry in the analysis of Thirring model in terms of SD equation. 
Kondo\cite{Thirring in inversion} further analyzed such a formulation 
in terms of the inversion method\cite{QCD in inversion}. 

It would be interesting to consider the critical behavior of the model 
in which the above two types of 4-fermi interaction coexist. 
Actually, Kim et al.\cite{korea} studied such a model but without manifest 
gauge symmetry. 
Due to each of the gauge symmetry their analysis of the SD equation 
unfortunately depends on their assumptions. 

In this paper, we analyze Gross-Neveu model with Thirring interaction 
by introducing hidden local symmetry 
and using the inversion method\cite{QCD in inversion} 
instead of SD equation. 
%
%
We obtained the generating functional of the next-to-leading order 
and the effective potential by the inversion method. 
The critical line is obtained in the parameter space $(g,g_v,N)$ 
where $g$ and $g_v$ are the dimensionless couplings 
of Gross-Neveu and Thirring model, respectively, 
and $N$ is the number of fermions. 
The phase structure drastically changes at g=1 
which is the critical coupling of the pure Gross-Neveu model 
at the leading order of $1/N$ expansion. 
For $g>1$ the phase structure is essentially determined 
by the Gross-Neveu interaction being only operative 
in the $1/N$-subleading corrections, 
while for $g<1$, the the symmetry breaking take place 
only due to the subleading effects where the Thirring interaction 
yields attractive force to form the condensate 
in contrast to the Gross-Neveu interaction which is opposite. 

Following the procedure of Ref.\cite{iksy}, 
we reformulate the original theory as a gauge theory 
in section \ref{hidden local symmetry}. 
In section \ref{invpot} and section \ref{critical}, 
we calculate the effective potential for the chiral order parameter 
$\langle \bar{\psi}\psi\rangle$, by using the inversion method 
in $1/N$ expansion. 
Based on this potential, we study spontaneous chiral symmetry 
breaking and explicitly obtain the critical surface which is 
the boundary between symmetric phase and spontaneously broken phase. 
Comparison with the SD equation is discussed in section \ref{sdkurabe}.

%
%
%
%
%
%
%
%
%
%
\section{Hidden Local Symmetry}
\label{hidden local symmetry}
\indent
%
%
%

Let us consider $D$-dimensional 
Gross-Neveu model with Thirring interaction $(2<D<4)$. 
The Lagrangian is given by 
\begin{equation}
\tilde{{\cal L}}=\bar{\Psi}(x)i\!\not{\!\partial}\Psi(x)
         +\frac{G_S}{2N}(\bar{\Psi}(x)\Psi(x))^2
         -\frac{G_V}{2N}(\bar{\Psi}(x)\gamma^\mu \Psi(x))^2
         +J\bar{\Psi}(x)\Psi(x),
\label{e:1}
\end{equation}
where $\Psi^a$ is a 4-component Dirac spinor 
with index $a$ $(a=1,2,\cdots,N)$ 
and belongs to a fundamental representation of $U(N)$, 
and $\gamma_\mu(\mu=0,1,2,\cdots,D-1)$ 
are $4\times 4$ matrices satisfying the Clifford algebra 
$\{ \gamma_\mu,\gamma_\nu \}=2g_{\mu\nu}{\bf 1}$. 
Index $a$ will be omitted in the following. 
Here we introduced the source $J=\mbox{constant}$ 
in order to study $\langle \bar{\Psi}\Psi \rangle$. 
Using the auxiliary field, 
we may rewrite this Lagrangian in to the form 
\begin{equation}
\tilde{{\cal L}}'=\bar{\Psi}(i\!\not{\!\partial}+J)\Psi
         -\frac{1}{2G_S}\sigma^2
         -\frac{1}{\sqrt{N}}\sigma\bar{\Psi}\Psi
         +\frac{1}{2G_V}\tilde{A}^2_\mu
         +\frac{1}{\sqrt{N}}\bar{\Psi}\tilde{\not{\!\!A}}\Psi,
\label{e:2}
\end{equation}
where the auxiliary massive vector field $\tilde{A_\mu}$ 
corresponding to the current $\bar{\Psi}\gamma_{\mu}\Psi$ 
is not really a gauge boson as it stands, 
and this Lagrangian actually has no gauge symmetry. 

Itoh et al.\cite{iksy} 
introduced hidden local symmetry to the Thirring model 
and treated it really as a gauge theory. 
Following the same procedure as in Ref.\cite{iksy}, 
we introduce hidden local symmetry to our model. 
What we should do is only the following substitution: 
%
\begin{equation}
\left\{
\begin{array}{ll}
 &\Psi=e^{-i\phi}\psi,\ \ \ \ \ \\
 &\tilde{A}_\mu=A_\mu-\sqrt{N}\partial_\mu \phi \ .\ \ \ \ \ 
\end{array}
\right.
\label{e:3}
\end{equation}
The Lagrangian(\ref{e:2}) is written as
\begin{equation}
\tilde{{\cal L}}''=\bar{\psi}(i\!\not{\!\partial}+J)\psi
         -\frac{1}{2G_S}\sigma^2
         -\frac{1}{\sqrt{N}}\sigma\bar{\psi}\psi
         +\frac{1}{2G_V}(A_\mu-\sqrt{N}\partial_\mu\phi)^2
         +\frac{1}{\sqrt{N}}\bar{\psi}\not{\!\!A}\psi\ ,
\label{e:4}
\end{equation}
where $\phi$ is the NG boson for $U(1)_{\mbox{local}}$ symmetry breaking, 
which is to be absorbed into the longitudinal mode of 
massless gauge boson $A_\mu$. 
$\tilde{{\cal L}}''$ is gauge-equivalent 
to the Lagrangian $\tilde{{\cal L}}'$. 
Actually, if we take the unitary gauge ($\theta=-\phi$), 
the Lagrangian $\tilde{{\cal L}}''$ reduces to $\tilde{{\cal L}}'$. 
It is noted that 
in contrast to $\tilde{{\cal L}}'$ which has no gauge symmetry, 
$\tilde{{\cal L}}''$ has a $U(1)_{local}$ symmetry: 
%
\begin{equation}
\left\{
\begin{array}{ll}
 &A_\mu\longrightarrow A'_\mu=A_\mu+\sqrt{N}\partial_\mu\theta\ ,\\
 &\psi\longrightarrow \psi'=e^{i\theta}\psi\ ,\\
 &\phi\longrightarrow \phi'=\phi+\theta\ .
\end{array}
\right.
\label{e:5}
\end{equation}
%
The total Lagrangian can be obtained 
by adding to (\ref{e:4}) the gauge fixing term 
and Faddeev-Popov(FP) ghost term: 
\begin{equation}
{\cal L}_{GF+FP}=-i\delta_B(\bar{c}f[A,c,\bar{c},B,\phi])\ ,
\label{e:6}
\end{equation}
and is invariant under the following BRS transformation: 
\begin{eqnarray}
\delta_B A_\mu(x)&=&\partial_\mu c(x)\ ,\nonumber\\
\delta_B B(x)&=&0\ ,\nonumber\\
\delta_B c(x)&=&0\ ,\nonumber\\
\delta_B \bar{c}(x)&=&iB(x)\ ,\nonumber\\
\delta_B \phi(x)&=&\frac{1}{\sqrt{N}}c(x)\ ,\nonumber\\
\delta_B \psi(x)&=&\frac{i}{\sqrt{N}}c(x)\psi(x)\ ,\nonumber\\
\delta_B \sigma(x)&=&0\ ,
\label{e:7}
\end{eqnarray}
where $c(x),\bar{c}(x)$ are the FP ghost fields 
and $B(x)$ is the Nakanishi-Lautrap field. 
If we chose $f(A,c,\bar{c},B,\phi)=F(A,\phi)+\frac{\xi}{2}B$, 
FP ghost is decoupled from the system. 
By integrating out $B$, 
we can rewrite gauge fixing term ${\cal L}_{GF}$ as 
\begin{equation}
{\cal L}_{GF}=-\frac{1}{2\xi}F(A,\phi)^2\ .
\label{e:8}
\end{equation}
Here, covariant gauge and $R_\xi$ gauge corresponds to 
following choice for $F(A,\phi)$, respectively. 
\begin{equation}
F(A,\phi)=
\partial^\mu A_\mu :\mbox{covariant gauge}\ ,
\label{e:9.1}
\end{equation}
\begin{equation}
F(A,\phi)=
\partial^\mu A_\mu+\frac{\sqrt{N}}{G_V}\xi\phi:R_\xi \mbox{ gauge}\ .
\label{e:9.2}
\end{equation}
In the covariant gauge(\ref{e:9.1}), NG boson $\phi(x)$ is not decoupled 
except the Landau gauge$(\xi=0)$. 
On the other hand, in the case of $R_{\xi}$ gauge, 
$\phi(x)$ is completely decoupled from the system 
and therefore the system can easily be treated\cite{iksy}. 
Here we choose $R_{\xi}$ gauge. 
The total Lagrangian is obtained by adding ${\cal L}_{GF}$ 
to the Lagrangian $\tilde{{\cal L}}''$: 
\begin{equation}
\tilde{{\cal L}}'''={\cal L}+{\cal L}_\phi\ ,
\label{e:10}
\end{equation}
\begin{equation}
{\cal L}_\phi=\frac{N}{G_V}\left\{  
\frac12 (\partial\phi)^2-\frac{\xi}{2G_V}\phi^2
  \right\}\ ,
\label{e:11}
\end{equation}
\begin{equation}
{\cal L}=\bar{\psi}(i\!\not{\!\partial}+J)\psi
         +\frac{1}{\sqrt{N}}\bar{\psi}\not{\!\!A}\psi
         -\frac{1}{\sqrt{N}}\sigma\bar{\psi}\psi
         -\frac{1}{2G_S}\sigma^2
         +\frac{1}{2G_V}A_\mu^2
         -\frac{1}{2\xi}(\partial^\mu A_\mu)^2\ ,
\label{e:12}
\end{equation}
\begin{equation}
{\cal L}=\bar{\psi}(i\!\not{\!\partial}+J)\psi
          +\frac12 \sigma iG^{(0)-1}_\sigma\sigma
          +\frac12 A^\mu iD^{(0)-1}_{\mu\nu}A^\nu
          +\frac{1}{\sqrt{N}}\bar{\psi}\not{\!\!A}\psi
          -\frac{1}{\sqrt{N}}\sigma\bar{\psi}\psi\ ,
\label{e:13}
\end{equation}
where $G^{(0)}_{\sigma}$ and $D^{(0)}_{\mu\nu}$ 
are the tree level scalar propagator 
and the tree level gauge boson propagator, respectively, 
which are written as 
\begin{eqnarray}
G^{(0)}_{\sigma}&=&\frac{i}{-G_S} \ ,\\
\label{e:13.1}
D^{(0)}_{\mu\nu}
&=&i\left\{
  \frac{1}{G_V^{-1}}P^T_{\mu\nu}
  +\left(  \frac{1}{G_V^{-1}+\xi^{-1}p^2}P^L_{\mu\nu}  \right)
 \right\}\ ,\\
\label{e:13.2}
P^T_{\mu\nu}&\equiv &g_{\mu\nu}-\frac{p_\mu p_\nu}{p^2}
\ ,\ 
P^L_{\mu\nu}\equiv \frac{p_\mu p_\nu}{p^2}\ .
\label{e:13.3}
\end{eqnarray}
%
%
%
%
%
%
\section{Vacuum Polarization}
\label{vacuum polarization}
\indent
%
%
%

In this section, 
we calculate the vacuum polarization tensor in the Euclidean space. 
By making use of the gauge invariant Pauli-Villars regularization, 
it is shown that the 1-loop vacuum polarization tensor has 
the following form\cite{iz qft}: 
%
\begin{equation}
\Pi_{\mu\nu}(p)=\Pi(p)P^T_{\mu\nu}\ ,
\label{e:18}
\end{equation}
%
where $\Pi(p)$ can be written as
\begin{eqnarray}
&&\Pi(p)=\frac{-2(tr1)\Gamma (2-D/2)}{(4\pi)^{D/2}}
         \int_0^1 dx \frac{(x-x^2)p^2}{\{J^2+(x-x^2)p^2 \}^{2-D/2}}\\
\label{e:19}
&&      =-\frac{(tr1)\Gamma (2-D/2)}{3(4\pi)^{D/2}}p^2J^{D-4}
         F\left( 2-\frac{D}{2},2,\frac52;-\frac{p^2}{4J^2} \right)\ ,
\label{e:20}
\end{eqnarray}
%
with $F(\alpha,\beta,\gamma;z)$ being the hypergeometric function. 
Since the source $J$ is infinitesimal, 
we can consider $J$ as $2J/p\ll 1$ for non-zero momentum $p\ne 0$. 
Using the mathematical identity for the hypergeometric function, 
%
\begin{eqnarray}
F(\alpha,\beta,\gamma;z)
&=&
\frac{\Gamma(\gamma)\Gamma(\beta-\alpha)}
     {\Gamma(\beta)\Gamma(\gamma-\alpha)}(-z)^{-\alpha}
        F(\alpha,\alpha-\gamma+1,\alpha-\beta+1;z^{-1})\nonumber\\
&+&
\frac{\Gamma(\gamma)\Gamma(\alpha-\beta)}
     {\Gamma(\alpha)\Gamma(\gamma-\beta)}(-z)^{-\beta}
        F(\beta,\beta-\gamma+1,\beta-\alpha+1;z^{-1})\ ,
\label{20.1}
\end{eqnarray}
we can expand $\Pi(p)$ in terms of $2J/p$: 
\begin{equation}
\Pi(p)=
   f_0p^{D-2}+f_2p^{D-4}J^{2}+f_Dp^{-2}J^{D}+f_4p^{D-6}J^{4}+O(J^6,J^{D+2})\ ,
\label{e:21}
\end{equation}
\begin{eqnarray}
&&f_0=\frac{-2(tr1)\Gamma (2-D/2)}{(4\pi)^{D/2}}B(D/2,D/2)\ ,\\
\label{e:22}
&&f_2=-\frac{2(4-D)(D-1)}{(D-2)}f_0\ ,\\
\label{e:23}
&&f_4=-\frac{2(6-D)(D-3)(D-1)}{(D-2)}f_0\ ,\\
\label{e:24}
&&f_{D}=\frac{2^3 f_0}{D(D-2)B(D/2,D/2)}\ .
\label{e:25}
\end{eqnarray}
%
%
%

Doing almost the same calculation as the one in $\Pi_{\mu\nu}(p)$, 
the 1-loop vacuum polarization for scalar propagator $\Pi_{\sigma}(p)$ 
is obtained as 
\begin{equation}
\Pi_{\sigma}(p)
= -c_0+c_1\int_0^1 dx \{ J^2+(x-x^2)p^2 \}^{D/2-1}\ ,
\label{e:27}
\end{equation}
where $c_0$ and $c_1$ are defined by
\begin{equation}
c_0=\frac{2(tr1)\Lambda^{D-2}}{(4\pi)^{D/2}\Gamma(D/2)(D-2)}\ ,\ 
c_1=\frac{2(tr1)\Gamma (2-D/2)(D-1)}{(4\pi)^{D/2}(D-2)}\ .
\label{e:26}
\end{equation}
%
For $p=0$, 
%
\begin{equation}
\Pi_{\sigma}(0)=-c_0+c_1J^{D-2}\ .
\label{e:27.1}
\end{equation}
For $p\ne 0$, $\Pi_{\sigma}$ can be expanded 
under the condition $2J/p\ll 1$: 
\begin{equation}
\Pi_\sigma(p)=
    -c_0+h_0p^{D-2}+h_2p^{D-4}J^{2}+h_Dp^{-2}J^{D}+h_4p^{D-6}J^{4}\ ,
\label{e:28}
\end{equation}
\begin{eqnarray}
&&h_0=\frac{(tr1)\Gamma(D/2)\Gamma(2-D/2)(D-1)}{(4\pi)^{D/2}2^{D-2}(D-2)}
      \ ,\\
\label{e:29}
&&h_2=2(D-1)h_0\ ,\\
\label{e:30}
&&h_4=2(D-3)(D-1)h_0\ ,\\
\label{e:31}
&&h_{D}=-\frac{(tr1)\Gamma(2-D/2)\Gamma(D/2)(D-1)}
              {(4\pi)^{D/2}\Gamma(D)D2^{1-D}(D-2)}\ .
\label{e:32}
\end{eqnarray}
%
When $2J/p>1$, $\Pi_{\sigma}(p)$ should be expanded in terms of $p/2J$. 
However it has already been shown in the Thirring model 
that the contribution 
from the region $2J/p>1$ is sufficiently small 
and therefore does not affect their conclusion
\cite{Thirring in inversion}. 
Since essentially the same analysis can be applied to our case, 
we use (\ref{e:28}) as long as $p\ne 0$. 
%
%
%
%
%
\section{Generating Functional}
\label{generating}
\indent
%
%
%
%

The generating functional $W[J]$ is given by 
\begin{equation}
e^{iW[J]}=
\int{\cal D}\psi{\cal D}\bar{\psi}{\cal D}\sigma{\cal D}Ae^{iS}\ .\\
\label{e:14}
\end{equation}
%
%
First we integrate out the fermion fields: 
%
\begin{eqnarray}
e^{iW[J]}
  &=&  \int{\cal D}\sigma{\cal D}A\mbox{Det}
         \left\{  i\!\not{\!\partial}-\frac{1}{\sqrt{N}}\sigma
                   +\frac{1}{\sqrt{N}}\not{\!\!A}
         \right\} \nonumber\\
 && \hspace{4em}
      \times \exp \left\{
          \frac12 \sigma iG^{(0)-1}_{\sigma}\sigma
          +\frac12 A^\mu iD^{(0)-1}_{\mu\nu}A^\nu 
       \right\}\\
\label{e:15}
 &\equiv &  \int{\cal D}\sigma{\cal D}Ae^{iS_{eff}}\ .
\label{e:16}
\end{eqnarray}
%
%
$S_{eff}$ is the effective action which includes fermion loop effects: 
\[
\begin{array}{ll}
iS_{eff}[\sigma,A]
&\hspace*{21em}\\
=
i\left\{
  \frac12 \sigma iG^{-1}_{\sigma} \sigma 
+ \frac12 A^\mu iD^{-1}_{\mu \nu} A^\nu
\right\}
&\hspace*{21em}
\end{array}
\]
%
%
%
\begin{center}
\begin{picture}(0,0)%
\epsfbox{svfig1-1.eps}
\end{picture}%
\setlength{\unitlength}{0.0125in}%
\begin{picture}(335,65)(40,747)
\end{picture}
\end{center}
\begin{center}
\begin{picture}(0,0)%
\epsfbox{svfig1-2.eps}
\end{picture}%
\setlength{\unitlength}{0.0125in}%
\begin{picture}(297,60)(30,763)
\end{picture}
\end{center}
\begin{equation}
\label{e:17}
\end{equation}
%
%
%
%
where $G_{\sigma}(p)$ and $D_{\mu\nu}(p)$ are 
the scalar propagator the gauge propagator, respectively, 
at the leading order of the $1/N$ expansion: 
%
\begin{eqnarray}
iG_{\sigma}^{-1}(p)&=&iG_{\sigma}^{(0)-1}(p)-\Pi_{\sigma}(p)\ ,\\
\label{e:17.01}
iD^{-1}_{\mu\nu}(p)&=&iD^{(0)-1}_{\mu\nu}(p)-\Pi_{\mu\nu}(p)\ .
\label{e:17.02}
\end{eqnarray}
After integrating out the fermion field, 
we have an infinite number of non-local vertices in eq.(\ref{e:17}). 
$W[J]$ is given by calculating all vacuum graphs 
which are made by connecting these non-local vertices 
in the $1/N$ expansion. 
%
%
%
\label{jbeki}

First we restrict ourselves 
to the leading order of $1/N$ expansion, $O(N)$. 
Since the source $J$ is infinitesimal, the higher order terms 
with respect to $J$ can be omitted in the calculation of $W[J]$. 
We expand $W[J]$ with respect to $J$ 
and we will omit the terms higher than $J^D$. 
Therefore we consider the order of various graphs 
in order to estimate which graphs should be included 
in $W[J]$ up to $O(J^{D})$.
%
For example, we estimate the graphs Fig.1-2 which are contained 
in the order parameter 
$(\varphi(J)=\langle \bar{\psi}\psi\rangle_{J})$: 
%
%
%
%
%
%
%
%
%
%
\begin{center}
\begin{picture}(0,0)%
\epsfbox{svfig5.eps}
\end{picture}%
\setlength{\unitlength}{0.0125in}%
\begin{picture}(233,139)(44,648)
\end{picture}
\end{center}
We have 
%
\begin{eqnarray}
\mbox{Fig.1}&=&\langle \bar{\psi}\psi\rangle_{0J}
=-iNtr\int_p\frac{i}{\not{p}+J}\nonumber\\
&=&Nc_0J-\frac{Nc_1}{D-1}J^{D-1} \ .
\label{jbeki:1}
\end{eqnarray}
%
%
This graph is of order $O(J)$. 
From (\ref{jbeki:1}), 
we also note that $1/N$ leading tadpole diagram Fig.2 
is of order $O(J)$. 

Let us consider the $n$-point vertex shown in Fig.3 ($n\geq 3$). 
The case $n=2$ in Fig.3 is the vacuum polarization 
and has already been absorbed into $G_\sigma$. 
This vertex is obtained by differentiating (\ref{jbeki:1}) 
with respect to source $J$: 
%
%
%
%
%
%
%
%
\begin{center}
\begin{picture}(0,0)%
\epsfbox{svfig6.eps}
\end{picture}%
\setlength{\unitlength}{0.0125in}%
\begin{picture}(469,175)(43,645)
\end{picture}
\end{center}
%
%
%
%
\begin{eqnarray}
\mbox{Fig.3}
&=&
\frac{1}{(n-1)!}
\left(\frac{\partial}{i\partial J}\right)^{n-1}
\left[   Nc_0J-\frac{Nc_1}{D-1}J^{D-1}   \right]
\nonumber\\
&=&
\frac{-Nc_1(D-1)(D-2)\cdots(D-n+1)}{i^{n-1}(n-1)!(D-1)}J^{D-n}
\label{jbeki:3.1}
\\
&\sim&O(J^{D-n})\ .
\label{jbeki:3}
\end{eqnarray}
%
%
%
If $n>D$, 
(\ref{jbeki:3.1}) has negative power in $J$ and diverges when $J=0$. 
This is the consequence of existence 
of infrared divergence of Fig.3 in the case of $J=0$. 
By using this vertex, 
we find that there exists a graph contributing to $W[J]$ at $O(J^D)$. 
Since Fig.4 includes Fig.3 ($O(J^{D-n})$) and $n$ tadpoles $(O(J^n))$, 
Fig.4 becomes of order $O(J^{D})$: 
%
%
\begin{equation}
\mbox{Fig.4}\sim O(J^{D})\ .
\end{equation}
%

It will be shown below that a graph 
which contains more than two non-local vertices of Fig.3 
does not contribute to $W[J]$. 
It is clear	
that the tadpole diagram Fig.5 is of order $O(J^{D-1})$ with $D>2$, 
and hence the order of this graph is higher 
than the simple tadpole Fig.2. 
Let us consider a graph which contains more than two vertices in Fig.3. 
There is at least one tadpole Fig.5. 
If we convert this tadpole Fig.5 into a simple tadpole Fig.2, 
the order of the graph becomes lower, 
and finally we can reduce the graph to Fig.4. 
Therefore we conclude that the order of such a graph 
as contains more than two vertices in Fig.3 is higher than that of Fig.4 
which is $O(J^{D})$, 
and such a graph does not contribute to this order. 
We can easily found here all graphs Fig.6-8 contributing to $W[J]$ 
at $O(J^D)$ at the leading order of $1/N$ expansion: 
%
%
%
%
%
%
%
%
\begin{center}
\begin{picture}(0,0)%
\epsfbox{svfig7.eps}
\end{picture}%
\setlength{\unitlength}{0.0125in}%
\begin{picture}(409,154)(21,665)
\end{picture}
\end{center}
%
%
These graphs are easily calculated 
by using (\ref{jbeki:1}) and (\ref{jbeki:3.1}). 

Similar arguments can be applied to 
the next-to-leading order of $1/N$ expansion. 
In the next-to-leading order, 
we have to consider the following non-local vertices in Fig.9-14 
similar to Fig.3. 
%
%
%
%
%
%
%
%
%
\begin{center}
\begin{picture}(0,0)%
\epsfbox{svfig8.eps}
\end{picture}%
\setlength{\unitlength}{0.0125in}%
\begin{picture}(374,151)(42,655)
\end{picture}
\end{center}
\begin{center}
\begin{picture}(0,0)%
\epsfbox{svfig9.eps}
\end{picture}%
\setlength{\unitlength}{0.0125in}%
\begin{picture}(394,154)(22,665)
\end{picture}
\end{center}
Fig.9 can be calculated by differentiating 
the scalar vacuum polarization (\ref{e:28}). 
\begin{eqnarray}
\mbox{Fig.9}
&=&
\frac12 
Tr\left[  G_\sigma 
            \frac{\partial}{i\partial J}(-i\Pi_\sigma)  
\right]
\\
\label{jbeki:4}
&=&
\frac{-\Omega}{2}\int_p
 \left[  iG_\sigma(p) 
            \frac{\partial}{\partial J}\Pi_\sigma(p)
\right]\ ,
\end{eqnarray}
%
%
where $\Omega$ denotes the space-time volume, $\Omega=\int d^{D}x$. 
$G_\sigma(p)$ and $\Pi_{\sigma}(p)$ can be expanded in terms of $J$. 
Then Fig.9 is given by 
\begin{eqnarray}
\mbox{Fig.9}
&=&
  \frac{-1}{2}\int_p
   \frac{1}{G_S^{-1}-c_0+h_0p^{D-2}}
   \left\{
    1-\frac{h_2p^{D-4}J^2}{G_S^{-1}-c_0+h_0p^{D-2}}
   \right\}
  \nonumber\\
  &&\hspace*{2em}
   \times \left\{  2h_2p^{D-4}J+Dh_Dp^{-2}J^{D-1}  \right\}
  +O(J^{D+1},J^{3})\ ,  \\
&\equiv&
  -B_1J-B_2J^{D-1}+O(J^3)\ ,\\
\label{jbeki:4.1}
&\sim& O(J)\ ,
\end{eqnarray}
%
where 
%
\begin{eqnarray}
&&B_1=\frac{4\alpha(D-1)\Lambda^{D-2}}{(D-2)}
      \left[ 1-\frac{1}{\tilde{g}_s}\log(1+\tilde{g}_s) \right]\ ,\\
\label{e:49}
&&B_2=\frac{\alpha Dh_D}{(D-2)h_0}\log(1+\tilde{g}_s)\ ,
\label{e:50}
\end{eqnarray}
\begin{equation}
g = c_0G_S\ ,\ 
\tilde{g}_s 
= \frac{h_0
\Lambda^{D-2}}{c_0}\left(\frac{g}{1-g}\right)\ ,\ 
\alpha=\frac{1}{(4\pi)^{D/2}\Gamma(D/2)}\ .
\label{e:50.1}
\end{equation}
Fig.10,11,12,13, and generally $n$($n>3$)-point vertex(Fig.14), 
can be obtained by the same procedure: 
%
\begin{equation}
\mbox{Fig.10}+\frac{1}{2}\mbox{Fig.11}=
\frac{1}{2\cdot 2}Tr
\left[ 
   G_\sigma \left(\frac{\partial}{i\partial J}\right)^2(-i\Pi_\sigma)
  \right]
\sim O(1)\ ,
\label{jbeki:5}
\end{equation}
\begin{equation}
\mbox{Fig.12}+\mbox{Fig.13}=
\frac{1}{2\cdot 3!} Tr
\left[
 G_\sigma \left(\frac{\partial}{i\partial J}\right)^3(-i\Pi_\sigma)
\right]
\sim O(J^{D-3})\ ,
\label{jbeki:6}
\end{equation}
\begin{eqnarray}
\mbox{Fig.14}
=
\frac{1}{2\cdot n!} Tr
\left[
 G_\sigma \left(\frac{\partial}{i\partial J}\right)^n(-i\Pi_\sigma)
\right]
\sim O(J^{D-n})\ .
\label{jbeki:7}
\end{eqnarray}
%
Thus $n(\geq 3)$-point vertex Fig.14 as well as Fig.3 
is of order $O(J^{D-n})$ 
and this fact leads to the same argument as at the leading order. 
Note that there appear several tadpoles($O(J)$) Fig.15-17 
in the next-to-leading order: 
%
%
%
%
%
%
%
%
\begin{center}
\begin{picture}(0,0)%
\epsfbox{svfig10.eps}%
\end{picture}%
\setlength{\unitlength}{0.0125in}%
\begin{picture}(355,141)(25,689)
\end{picture}
\end{center}
%
By using almost the same discussion as in the leading order, 
we find the following graphs will contribute to generating functional 
in the next-to-leading order. 
The above tadpoles Fig.15-17 and the following graphs Fig.18-24 
can be easily calculated by use of (\ref{jbeki:1}), 
(\ref{jbeki:4.1}), (\ref{jbeki:5}) and (\ref{jbeki:7}). 
%
%
%
%
%
%
%
\begin{center}
\begin{picture}(0,0)%
\epsfbox{svfig11.eps}%
\end{picture}%
\setlength{\unitlength}{0.0125in}%
\begin{picture}(428,167)(33,658)
\end{picture}
\end{center}
\begin{center}
\begin{picture}(0,0)%
\epsfbox{svfig12.eps}%
\end{picture}%
\setlength{\unitlength}{0.0125in}%
\begin{picture}(466,170)(62,610)
\end{picture}
\end{center}
%
%
%

Now we include the gauge interaction, 
which can be performed by almost the same procedure as before 
except that gauge tadpoles do not exist. 
Here we denote only graphs(Fig.25-31) 
contributing to the generating functional: 
%
%
%
%
%
\begin{center}
\begin{picture}(0,0)%
\epsfbox{svfig13.eps}%
\end{picture}%
\setlength{\unitlength}{0.0125in}%
\begin{picture}(419,171)(42,658)
\end{picture}
\end{center}
\begin{center}
\begin{picture}(0,0)%
\epsfbox{svfig14.eps}%
\end{picture}%
\setlength{\unitlength}{0.0125in}%
\begin{picture}(466,170)(62,610)
\end{picture}
\end{center}
%
%
In order to calculate these graphs, 
we need Fig.32 
which is obtained by differentiating the vacuum polarization tensor: 
\begin{center}
\begin{picture}(0,0)%
\epsfbox{svfig15.eps}%
\end{picture}%
\setlength{\unitlength}{0.0125in}%
\begin{picture}(94,138)(229,503)
\end{picture}
\end{center}
%
%
Namely, 
\begin{equation}
\mbox{Fig.32}=-A_1J-A_2J^{D-1}+O(J^3)\ ,
\end{equation}
\begin{eqnarray}
&&A_1=\frac{2\alpha (D-1)f_2\Lambda^{D-2}}{(D-2)f_0}
      \left[ 1-\frac{1}{g_V}\log(1+g_V) \right]\ ,\\
\label{e:47}
&&A_2=\frac{\alpha D(D-1)f_D}{(D-2)f_0}\log(1+g_V)\ ,
\label{e:48}
\end{eqnarray}
where $g_V\equiv -f_0\Lambda^{D-2}G_V$ is 
the dimensionless Thirring coupling constant. 

In the end of lengthy calculation, 
we found all graphs(Fig.6-8 and Fig.18-31) 
contributing to the generating functional $W[J]$ 
in the next-to-leading order of $1/N$ expansion. 
By calculating all these graphs, we finally obtain $W[J]$ as 
\begin{eqnarray}
\frac{1}{\Omega}W[J]
&=&
\mbox{(Fig.6)}+\mbox{(Fig.7)}+\mbox{(Fig.8)}
+ \mbox{(Fig.18)}+ \cdots +\mbox{(Fig.31)}\\
\label{jbeki:8}
&=&
\frac{1}{2}KJ^2 + \frac{1}{D}PJ^D \ ,
\label{jbeki:8.1}
\end{eqnarray}
where $K,P$ are given by 
\begin{eqnarray}
&&K=NK_0+K_1\ ,\\
\label{e:41}
&&K_0=\frac{c_0}{1-g}\ ,\\
\label{e:42}
&&K_1=-\frac{A_1+B_1}{(1-g)^2}\ ,\\
\label{e:43}
&&P=NP_0+P_1\ ,\\
\label{e:44}
&&P_0=\frac{-c_1}{D-1}\left(\frac{1}{1-g}\right)^{D}\ ,\\
\label{e:45}
&&P_1=\left\{
-A_2-B_2+\frac{c_1D(A_1+B_1)}{c_0(D-1)}\left(\frac{g}{1-g}\right)
\right\} \left(\frac{1}{1-g}\right)^{D}\ .
\label{e:46}
\end{eqnarray}
%
%
%
%
%
%
\section{Effective Potential by Inversion}
\label{invpot}
\indent

The order parameter for the dynamical symmetry breaking, $\varphi(J)$, 
is given by differentiating (\ref{jbeki:8.1}) with respect to source $J$: 
\begin{equation}
\varphi(J)
=\langle\bar{\psi}\psi\rangle_J
=\frac{1}{\Omega}\frac{\partial W}{\partial J}=KJ+PJ^{D-1}\  .
\label{e:35}
\end{equation}
We here made use of inversion method\cite{QCD in inversion} 
to obtain the effective potential 
instead of taking a Legendre transformation. 
By inverting (\ref{e:35}), we obtain 
\begin{eqnarray}
J
&=&
K^{-1}\varphi-K^{-1}PJ^{D-1}\\
\label{e:36}
&=&
K^{-1}\varphi-PK^{-D}\varphi^{D-1}\\
\label{e:37}
&=&
K^{-1}\varphi+Q\varphi^{D-1}\  .
\label{e:38}
\end{eqnarray}
$K^{-1}$ is expanded in $1/N$ as 
\begin{equation}
K^{-1}
=
\frac{1}{NK_0}\left( 1 - \frac{K_1}{NK_0} \right)\ .
\label{kinv}
\end{equation}
From the relation 
\begin{equation}
J
=
\frac{\partial V}{\partial \varphi}\ ,
\label{e:39}
\end{equation}
the effective potential of $D(2<D<4)$-dimensional 
Thirring-Gross-Neveu model can be obtained 
in the next-to-leading order of $1/N$ expansion: 
\begin{equation}
V
=
\frac12 K^{-1}\varphi^2+\frac{1}{D}Q\varphi^{D}\ .
\label{e:40}
\end{equation}
Dynamical symmetry breaking occurs 
if the equation $J=\frac{\partial V}{\partial\varphi}=0$ has 
a non-zero solution, $\varphi\ne 0$. 
It is realized when $K^{-1}<0$, 
and the critical surface is given by $K^{-1}=0$. 
Note that we calculate all graphs through vacuum polarization. 
Therefore our effective potential and all our conclusion 
has no gauge dependence\cite{QCD in inversion}. 
%
%
%
%
%
%
%
\section{Critical Line}
\label{critical}
\indent

$K^{-1}$ is given in (\ref{kinv}) reads 
\begin{eqnarray}
K^{-1}&=&\frac{1}{NK_0}\left(1-\frac{K_1}{NK_0}\right)\\
\label{e:51}
&=&\frac{1-g}{Nc_0}\left\{1+\frac{1-g}{Nc_0}\frac{(A_1+B_1)}{(1-g)^2}\right\}\\
\label{e:52}
&=&\frac{1}{Nc_0}\left\{ 1-g+\frac{A_1+B_1}{Nc_0} \right\}\ .
\label{e:53}
\end{eqnarray}
Now, the critical coupling constants are given 
by the equation $K^{-1}=0$ 
which defines a surface$(g_c,g_{V_c},N_c)$ 
in the parameter space $(g,g_V,N)$: 
\begin{equation}
0=1-g_c + \frac{(D-1)}{2N_c}
          \left[ 1-\frac{1}{\tilde{g}_{sc}}\log(1+\tilde{g}_{sc}) \right]
      - \frac{(D-1)^2(4-D)}{2N_c(D-2)}
          \left[ 1-\frac{1}{g_{Vc}}\log(1+g_{Vc}) \right]\ ,
\label{e:55}
\end{equation}
where $\tilde{g}_{sc}$ was defined in (\ref{e:50.1}).
We can see from (\ref{e:55}) that $g_c\sim 1$ for large $N_c$, 
and therefore 
we can approximate this equation as 
\begin{equation}
g_c
\simeq
 1 + \frac{D-1}{2N_c}
   - \frac{(D-1)^2(4-D)}{2N_c(D-2)}
     \left[ 1-\frac{1}{g_{Vc}}\log(1+g_{Vc}) \right]
  + O\left( \frac{1}{N_c^2}\log N_c \right)   \ .
\label{original}
\end{equation}
%
Eq.(\ref{original}) works well when $N$ is sufficiently large. 
We will investigate the structure of the critical surface 
using eq.(\ref{original}). 

Before investigating (\ref{original}) in detail, 
we consider the pure Gross-Neveu model. 
Taking $g_{Vc}=0$ in the eq.(\ref{original}), 
the critical line of the dynamical symmetry breaking 
in the Gross-Neveu model is given by 
\begin{equation}
g_c\simeq 1+\frac{D-1}{2N_c} +O\left( \frac{1}{N_c^2}\log N_c \right) \ .
\label{e:57}
\end{equation}
It is remarkable that this critical line is exactly the same line 
obtained by using SD equation up to the next-to-leading 
order of $1/N$ expansion\cite{wang}.


Let us return to (\ref{original}) to look into the critical surface 
of Thirring-Gross-Neveu model. 
We consider here $D=3$ for convenience. 
Then the critical line is written as 
\begin{equation}
g_c=1+\frac{1-2H(g_{Vc})}{N_c}\ ,
\label{e:101}
\end{equation}
where $H(z)$ is defined by 
\begin{equation}
H(z)
\equiv
1 - \frac{1}{z}\log (1+z)\ .
\end{equation}
%
%
$H(z)$ is monotonically increasing function and $H(0)=0, H(\infty)=1$. 
Since $0\leq H\leq 1$ and $N\geq 1$, we find $g_c <2$ in (\ref{e:101}). 
The critical line in the ($g_{V},N$) plane is given 
in FIG.A-1 and FIG.A-2 for various values of $g$. 

We here divide the region $0<g<2$ into three regions, 
(1)$g>2$, (2)$0<g<1$(FIG.A-1) and (3)$1<g<2$(FIG.A-2), 
where $g=1$ is the value of critical coupling constant 
at the leading order of the Gross-Neveu model. 
The forms of critical line in these region are different from 
each other as follows: 
\\
(1) $g>2$

In this case, the chiral symmetry is always broken 
irrespectively of $g_V$ and $N$. 
\\
(2) $1<g<2$ (FIG.A-1)

In this region, the critical line (\ref{e:101}) reads 
\begin{equation}
N_c
=
\frac{1-2H(g_{Vc})}{g_c-1}<\frac{1}{g_c-1} \ \ .
\label{e:102}
\end{equation}
For $g>1$, $g_{Vc}$ can take a value 
which satisfy the equation $2H(g_{Vc})<1$. 
The critical line exists only in the region $N\leq \frac{1}{g_c-1}$ 
and $N_c$ decreases as $g_{Vc}$ increases 
as shown in FIG.A-1. 
The symmetry is dynamically broken when $N>\frac{1}{g_c-1}$. 
\\
(3) $0<g<1$ (FIG.A-2)

In this case, $N_c$ increases according to increasing $g_{Vc}$ 
and there exists a critical value of $N_c$: 
\begin{equation}
N_c
= \frac{2H(g_{Vc})-1}{1-g_c} 
< N_c(g_c,g_{Vc}=\infty)=\frac{1}{1-g_c}\ .
\end{equation}
The dynamical symmetry breaking does not occur when $N>N_c(g_c,\infty)$. 
This critical line is quite similar to 
the critical line obtained in the Thirring model
\cite{iksy}\cite{Thirring in inversion}. 
Thirring interaction strongly affect the system in this region, $0<g<1$. 
Thus, the property of the critical line is drastically changed at $g=1$.

This can be understood naturally as follows. 
$g=1$ is the critical coupling constant of the Gross-Neveu model 
at the leading order of $1/N$ expansion, 
and Thirring interaction does not strongly affect the leading order. 
Therefore, for $g>1$, Gross-Neveu interaction is dominant. 
The system behaves like the Gross-Neveu model, 
and the symmetry is broken almost all the region. 
As $N$ increases, 
the effect of $1/N$ next-to-leading order becomes small, 
and the system is dominated by the tadpole of the leading order. 
Therefore the dynamical symmetry breaking occur 
in large $N$ region for $g>1$.

On the other hand, 
the critical behavior is quite different in $0<g<1$. 
In this region, the Thirring interaction strongly affects the system, 
and the system (or the critical line) becomes Thirring-like. 
There exists a critical value for $N$, $N_c(g_c,g_{Vc}=\infty)<\infty$. 
The dynamical symmetry breaking does not occur in the large $N$ region, 
$N>N_c(g_{Vc}=\infty,g_c)$, 
even if $g_V\longrightarrow\infty$. 
The symmetry is broken in the small $N$ region. 
This is in contrast to the case $1<g<2$ 
where the system behaves like Gross-Neveu model.

Thus the critical behavior drastically changes at $g=1$. 
The system behaves like Thirring model in $g<1$, 
whereas it does like Gross-Neveu model in $g>1$. 
%
%
%
%
%
%

\section{Conclusion and Discussion}
\label{sdkurabe}
\indent

  In this paper, we have constructed the effective potential 
for the order parameter of the chiral symmetry, 
the fermion condensate $\langle\bar{\psi}\psi\rangle$ 
in the hybrid model of the Gross Neveu model 
and the Thirring model in $D(2<D<4)$ dimensions. 
From this potential, we have shown existence of the chiral 
symmetry breaking and explicitly obtained the critical surface.

We have analysed our model using the inversion method in $1/N$ expansion. 
In this model, existence of Gross-Neveu interaction make 
our analysis complicated. 
So many diagrams contribute to the generating functional. 
But we found we do not have to calculate all the diagrams. 
We have expanded the generating functional $W[J]$ 
in terms of the infinitesimal source $J$, 
and estimate it on the order $O(J^{D})$. 
In section\ref{jbeki}, 
we investigated systematically the order of diagrams 
with respect to source $J$, 
and determine 
which diagram contributes to the generating functional $W[J]$ 
or effective potential $V(\langle\bar{\psi}\psi\rangle)$. 
Resultant effective potential has no gauge dependence, 
because we calculated all diagrams through vacuum polarization.

From this effective potential, we obtained explicitly the critical surface 
which separates symmetric phase and broken phase of the chiral symmetry. 
Especially, if we consider the special case $g_{V}=0$ (Gross-Neveu model), 
our result(\ref{e:57}) reproduces exactly the same result obtained 
by using SD equation\cite{wang}.


Kim et al.\cite{korea} have already studied Thirring Gross-Neveu model 
by using SD equation and they found critical surface. 
Although the original Lagrangian(\ref{e:2}) has no gauge symmetry, 
they pretended the auxiliary massive vector field as really a gauge field 
without introducing hidden local symmetry. 
Furthermore, 
the SD equation becomes coupled integral equations 
for $A(p)$ and $B(p)$ in the full fermion propagator 
$S(p)^{-1}=\not{\!p}A(p)-B(p)$, 
which are difficult to solve. 
Therefore additional approximations (or assumptions) 
besides $1/N$ expansion were made in their analysis. 
They assumed that the form of $A(p)$ can be determine 
by using perturbation, 
and mass function $B(p)$ can be regarded as a constant. 
As a result of such an approximation, 
the effects of the next-to-leading order of the Gross-Neveu interaction 
were not included.

On the other hand, 
we succeeded in obtaining the effective potential 
and the critical line without using additional 
approximation besides $1/N$ expansion.

\begin{center}
{\large Acknowledgments}
\end{center}

The author would like to thank 
Prof. K. Yamawaki, and Drs. T. Itoh and M. Sugiura 
for valuable discussion and comments.

%
%
%
%

%
%
%
%
%
\newpage
\epsfysize=15cm
\begin{center}
\leavevmode
\epsfbox{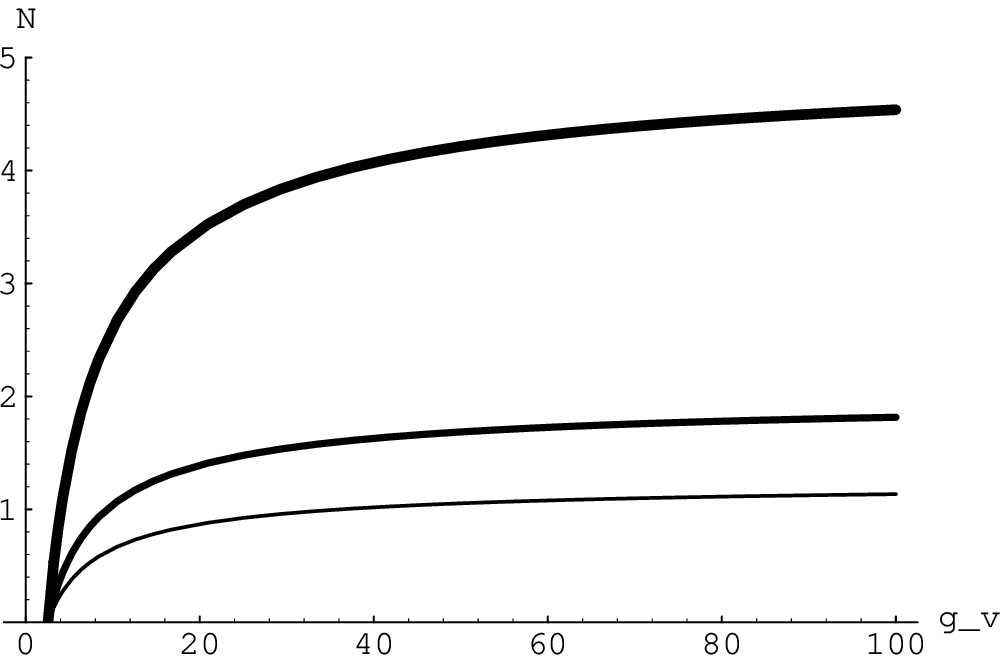}
\end{center}
Fig.A-1:\\
The critical lines for various values 
of Gross-Neveu coupling constant $g$, 
($g=0.2$: lower line), ($g=0.5$: middle line), ($g=0.8$: upper line). 
Horizontal line and vertical line denote $g_V$ and $N$, respectively.
\epsfysize=15cm
\begin{center}
\leavevmode
\epsfbox{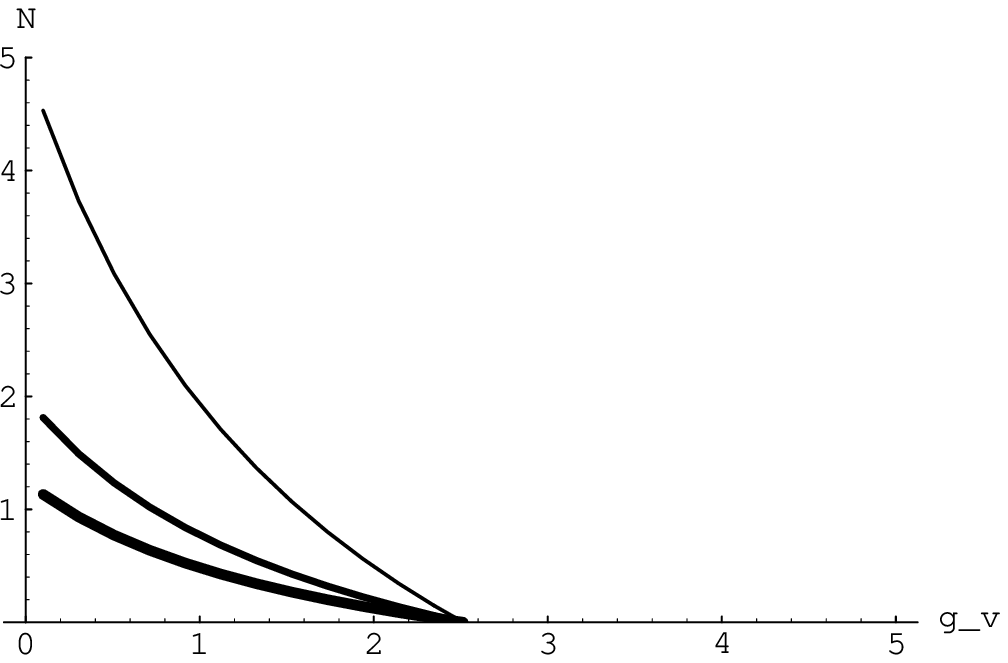}
\end{center}
Fig.A-2\\
The critical lines for various values 
of Gross-Neveu coupling constant $g$, 
($g=1.2$: upper line), ($g=1.5$: middle line), ($g=1.8$: lower line).
Horizontal line and vertical line denote $g_V$ and $N$, respectively. 

\end{document}